\documentclass[a4paper,11pt]{article}
\usepackage{pos}
\usepackage[backend=bibtex, sorting=none, bibstyle=ieee]{biblatex}  
\usepackage{placeins}
\usepackage{slashed}
\title{Calculating $\Delta m_K$ with lattice QCD}

\author*[\dag]{Bigeng Wang}
\notes{\note{This work was partially supported by US DOE grant
\#DE-SC0011941 and
used computer time provided by the Innovative and Novel Computational Impact
on Theory and Experiment (INCITE) program. This research used resources of
the Argonne Leadership Computing Facility, which is a DOE Office of Science
User Facility supported under Contract DE-AC02-06CH11357.}}

\affiliation{Department of Physics, Columbia University, New York, NY 10027, USA\\}

\emailAdd{bw2482@columbia.edu}

\abstract{We have completed a lattice QCD calculation of $\Delta m_K$, the mass difference between the long- and short-lived K mesons.  The calculation was performed on a $64^3 \times 128$ lattice using 152 configurations with physical quark masses and an inverse lattice spacing of $1/a=2.36$ GeV.   While the statistical error approaches a relatively small size of 9\%, several sources of systematic errors may have more significant effects. In this paper we will address studies performed on smaller lattices to estimate the systematic errors in our result. }

\FullConference{%
 The 38th International Symposium on Lattice Field Theory, LATTICE2021
  26th-30th July, 2021
  Zoom/Gather@Massachusetts Institute of Technology
}

\bibliography{ skeleton}

\AtEveryBibitem{\clearfield{issn}}
\AtEveryBibitem{\clearlist{issn}}

\AtEveryBibitem{\clearfield{language}}
\AtEveryBibitem{\clearlist{language}}

\AtEveryBibitem{\clearfield{doi}}
\AtEveryBibitem{\clearlist{doi}}

\AtEveryBibitem{\clearfield{url}}
\AtEveryBibitem{\clearlist{url}}

\AtEveryBibitem{%
  \ifentrytype{online}
    {}
    {\clearfield{urlyear}\clearfield{urlmonth}\clearfield{urlday}}}

\begin{document}
\maketitle

\section{Introduction}
The mass difference between the long- and short-lived K mesons, $\Delta m_K$, is generated by K meson mixing through $\Delta S = 2$ weak interaction 
and is closely related to the indirect CP violation parameter $\epsilon_K$. This tiny quantity has been precisely measured experimentally to be $3.484(6) \times 10^{-12}$ MeV \cite{PDG_2020} and the comparison between the prediction for this quantity by the standard model and its experimental value will serve as a detector of new physics beyond the standard model. The calculation has been extended from the first exploratory calculation with only connected diagrams to full calculations on near-physical\cite{klks_zbai_PhysRevLett.113.112003} and physical ensembles\cite{zbai_EPJ_klks}.

\section{Non-perturbative calculation of \texorpdfstring{$\Delta m_K$} using a renormalization scale above the charm quark mass}
\label{sec:dmk_formalism}
Due to the Glashow–Iliopoulos–Maiani(GIM) mechanism, the dominant contribution to $\Delta m_K$ comes from the charm quark scale and below and the calculation can be better performed by making the division of long and short distances at an energy scale larger than the charm mass and treating the charm quark non-perturbatively by using two $\Delta S =1$ operators in a lattice calculation. The $K_L-K_S$ mass difference is expressed as:
\begin{equation}
    \label{eqn:dmk}
    \Delta M_K = 2 \mathrm{Re} M_{\overline{0}0} = 2 \mathcal{P} \sum_{n} \frac{\langle \overline{K^0}|H_W|n\rangle\langle n|H_W|K^0\rangle}{m_K-E_n},
\end{equation}
where $H_W$ is the $\Delta S =1 $ effective Hamiltonian:
\begin{equation}
\label{eqn:hamiltonian}
    H_W = \frac{G_F}{\sqrt{2}} \sum_{q,q'=u,c} V_{qd}V^*_{q's}
    (C_1Q_1^{qq'}+C_2Q_2^{qq'}).
\end{equation}

\begin{figure}[htbp]
    \centering
    \includegraphics[width=0.8\textwidth]{ 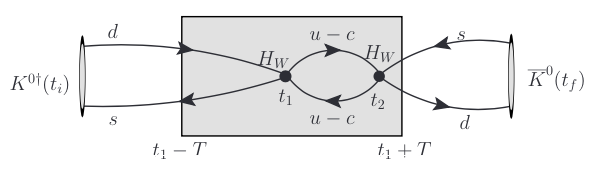}
    \caption{The single integration method on the lattice. The shadowed box refers to the region of integration.}
    \label{fig:single_int_method_original}
    \end{figure}

To calculate $\Delta m_K$  we can integrate the four-point correlation functions over the time locations of one of the weak operators with the other one being fixed as shown in Figure \ref{fig:single_int_method_original} and obtain the single-integrated correlator:
\begin{equation}
    \mathcal{A}^S (T)= \frac{1}{2!} \sum^{t_1+T}_{t_2=t_1-T} 
    \langle 0 |T \{ \overline{K^0}(t_f)H_W(t_2)H_W(t_1)K^0(t_i) \} | 0 \rangle.
    \label{eqn:single_intgrate}
    \end{equation}

Details about the method and the calculations with physical quark masses can be found in Reference \cite{Lattice_2018_Wang:2019vZ} and \cite{Lattice_2019_Wang:2020h0}. We have performed a calculation on an ensemble of 2+1 flavor gauge configurations with $a^{-1}=2.36$GeV and a $64^3 \times 128$ lattice volume using 152 configurations. Our preliminary result for $\Delta m_K$ is:
        \begin{equation}
            \Delta m_K  = 5.8(0.6)_{\mathrm{stat}} \times 10^{-12} \mathrm{MeV}.
        \end{equation}
While the statistical error approaches a relatively small size of 9\%, several sources of systematic errors may have more significant effects.

\section{Systematic errors}
Two potentially important systematic errors come from finite-volume and finite lattice spacing effects. The finite-volume correction to $\Delta m_K$ based on the formula proposed in Reference \cite{finite_volume_klks_PhysRevD.91.114510} is estimated to be:
    $\Delta m_K^{FV} = -0.22(7) \times 10^{-12}$ MeV.
As for the finite lattice spacing effects, the $\mathcal{O}(a^2)$ error due to the heavy charm is estimated to be the largest source of systematic error.
\subsection{Sources of \texorpdfstring{$\mathcal{O}(a^2)$} finite lattice spacing errors}
After eliminating the $\mathcal{O}(a)$ finite lattice spacing errors by our choice of fermion action, we have to estimate the remaining $\mathcal{O}(a^2)$ finite lattice spacing errors. The first possible source is associated with the heavy charm quark we have included in our lattice calculation. The effect should be proportional to the dimensionless quantity $(m_ca)^2$. Determination of the size of this term will allow us to estimate its contribution to our systematic error. 
%
%
%
\subsection{Scaling test on lattices with different lattice spacings}
A scaling test, in which we measure the same physical quantities on several lattices with different lattice spacings, can help us determine the size of the $\mathcal{O}(a^2)$ finite lattice spacing error and give an estimate of how large these effects are.

Thus, in order to estimate the finite lattice spacing errors for our $\Delta m_K$ calculation, we perform scaling tests focusing on the matrix elements obtained from three-point correlation functions and four-point integrated correlators using two different lattice spacings. It's economical to start with a smaller lattice where the relatively large $m_ca$ value is examined. We perform the scaling tests on the 24I and 32I ensembles and details about these two ensembles are listed in Table \ref{tab:dmk_lattice_info}.

\begin{table}[htbp]
    \centering
    \begin{tabular}{|c|c|c|c|c|c|c|c|c|} \hline
        Lattice & Action & $a^{-1}$& Lattice &$\beta$ & b+c& $L_s$  & $m_l$ & $m_h$ \\ 
        name  & (F+G) & (GeV) & Volume &  &    &   &  &  \\\hline
        24I &  DWF+I & 1.785(5)& $24^3 \times 64 \times 16$ & 2.13 & 1.0& 16 & 0.0050 & 0.0400  \\ \hline
        32I & DWF+I & 2.383(9)& $32^3 \times 64 \times 16$ & 2.25 & 1.0& 16  & 0.0040 & 0.0300  \\ \hline
        64I & MDWF+I &  2.359(7)& $64^3 \times 128 \times 12$ & 2.25 & 2.0& 12  & 0.000678 & 0.02661 \\ \hline
        32IF & DWF+I & 3.15(2)& $32^3 \times 64 \times 12$ & 2.37 & 1.0& 12  & 0.0047 & 0.0186 \\ \hline
    \end{tabular}
    \caption{Dynamical 2+1 flavor domain wall fermion lattices used in our $\Delta m_K$ calculation \cite{DWF_2014_PhysRevD.93.074505}. The fermion and gauge (F+G) action abbreviations are: DWF = domain wall fermions, MDWF = Mobius domain wall fermions, I = Iwasaki gauge action. $m_{l/h}$ are the light and heavy sea quark masses in lattice units.}
    \label{tab:dmk_lattice_info}
\end{table}

To obtain the input valence quark masses for the two ensembles which result in physical meson masses on the two lattices which are consistent, we first set the physical values of meson masses to be the ones obtained from the calculation on the 32IF ensemble using its unitary quark masses. Then based on several meson masses obtained on the 24I and 32I ensembles for various valence quark masses\cite{DWF_2014_PhysRevD.93.074505}, we perform interpolations to obtain the valence quark masses which yield physical meson masses consistent with the 32IF meson masses described above using formulas from chiral effective theory. The calculated valence quark masses and the expected meson masses are shown in Table \ref{tab:scaling_lattice_valence_input}.

\begin{table}[!htbp]
    \centering
    \begin{tabular}{|c|c|c|c|c|c|c|} \hline
        Lattice  &  $m_x$ & $m_y$ & $m_{\pi,\mathrm{pre}}a$ & $m_{\pi,\mathrm{pre}}$/MeV & $m_{K,\mathrm{pre}}a$ & $m_{K,\mathrm{pre}}$/MeV \\ \hline
        24I & 0.00667 & 0.0321 & 0.2079& 371.15 & 0.3125 & 557.83 \\ \hline
        32I & 0.00649 & 0.0249  & 0.1557 & 371.15 & 0.2332& 557.83 \\ \hline \hline
        Lattice  &  $m_{x,\mathrm{uni}}$ & $m_{y,\mathrm{uni}}$ & $m_{\pi,\mathrm{uni}}a$ & $m_{\pi,\mathrm{uni}}$/MeV & $m_{K,\mathrm{uni}}a$ & $m_{K,\mathrm{uni}}$/MeV \\ \hline
        32IF & 0.0047 & 0.0186 & 0.1179 & 371.15  & 0.1772 & 557.83  \\ \hline
    \end{tabular}
    \caption{Parameters related to the lattices for measurements. $m_x$ is the valence mass for light quarks: up and down. $m_y$ is the valence mass for strange quark. The predicted pion mass $m_{\pi,\mathrm{pre}}$ and the predicted kaon mass $m_{K,\mathrm{pre}}$ are displayed both in lattice units and in physical units.}
    \label{tab:scaling_lattice_valence_input}
\end{table}
\subsection{Results from two-point functions}
We expect our input valence quark masses to produce mesons with equal physical masses on the two different lattices. The results listed in Table \ref{tab:scaling_meson_masses} and Table \ref{tab:scaling_mD_mass_high_mc} verify that not only light mesons like pion and kaon, but also heavy charmed mesons with relatively large values of $m_c$, have consistent masses. We can therefore conclude the quantities we have calculated on these two ensembles with different lattice spacings are consistent in physics.

\begin{table}[!htbp]
    \centering
    \begin{tabular}{|c|c|c|c|c|c|c|c|} \hline
        Lattice  & $N_{\mathrm{conf}}$& $m_{\pi}$/MeV &$m_{\pi}a$ & $m_{\pi,\mathrm{pre}}a$ & $m_{K}$/MeV & $m_{K}a$ & $m_{K,pre}a$\\ \hline
        24I & 186 & 371.3(7) & 0.2080(4) &0.2079 & 556.2(7) &0.3116(4) & 0.3125 
        \\ \hline
        32I & 222 & 371.4(6) & 0.1558(2) &0.1557 & 557.5(6) &0.2340(3) & 0.2332  
        \\ \hline

    \end{tabular}
    \caption{The light meson masses resulting from light and heavy quark masses obtained from interpolation calculations on the 32I and 24I ensembles.}
    \label{tab:scaling_meson_masses}
    \end{table}

\begin{table}[!htbp]
        \centering
        \resizebox{\textwidth}{!}{
        \begin{tabular}{|c|c|c|c|c|c|c|c|} \hline
         $m_c$ 24I& $m_c$/GeV 24I & $m_c^R$/GeV 24I  & $m_D$ 24I/GeV & $m_c$ 32I & $m_c$/GeV 32I & $m_c^R$/GeV 32I  & $m_D$ 32I/GeV \\ \hline
         0.15&0.26775 &0.4079& 1.0891(27)& 0.11   & 0.26775 &0.4068&  1.1151(9)\\ \hline
         0.20&0.357 &0.5439& 1.2599(31)& 0.15   &0.357 &0.5423&  1.2940(10)\\ \hline
         0.25& 0.44625 &0.6799& 1.4142(37)& 0.19   &0.44625 &0.6779&  1.4563(11)\\ \hline
         0.30&0.5355 &0.8158& 1.5550(43)& 0.22   &0.5355 &0.8135&  1.6057(11)\\ \hline
         0.35&0.62475 &0.9518& 1.6836(50)& 0.26   &0.62475 &0.9491&  1.7442(12)\\ \hline 
        \end{tabular}
        }
        \caption{$m_c$ masses and corresponding D meson mass $m_D$ for the 24I and 32I ensembles, with renormalized masses using mass renormalization factors $Z_{m,\mathrm{24I}}^{\gamma}=1.5235(13)$ and $Z_{m,\mathrm{32I}}^{\gamma}=1.5192(39)$\cite{DWF_2014_PhysRevD.93.074505}. }.
        \label{tab:scaling_mD_mass_high_mc}
    \end{table}
\subsection{Scaling of three-point matrix elements}
\begin{figure}
        \centering
        \includegraphics[width=0.45\textwidth]{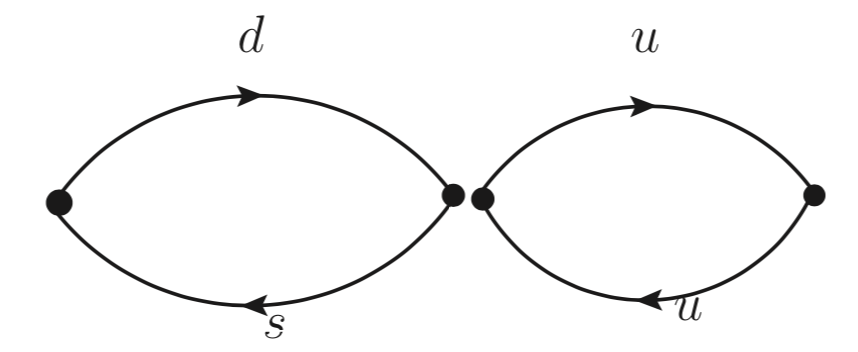} \hfill
        \includegraphics[width=0.45\textwidth]{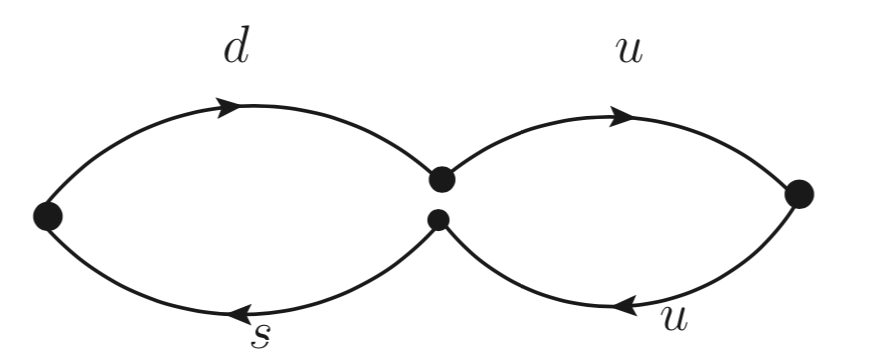} \hfill
        
        \centering
        \includegraphics[width=0.45\textwidth]{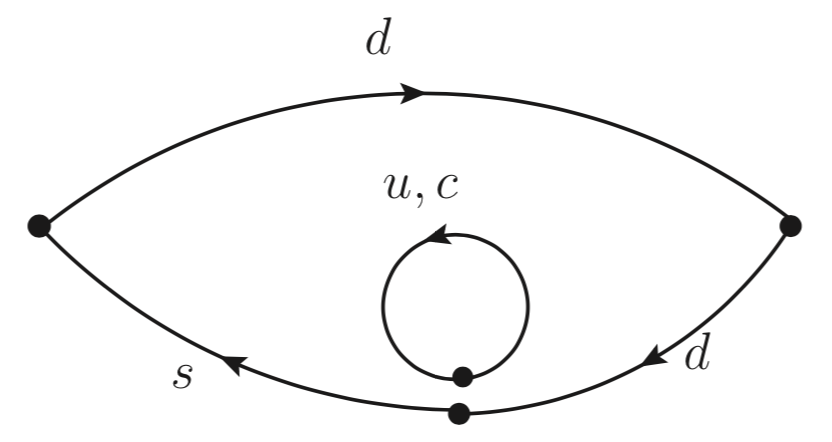} \hfill 
        \includegraphics[width=0.45\textwidth]{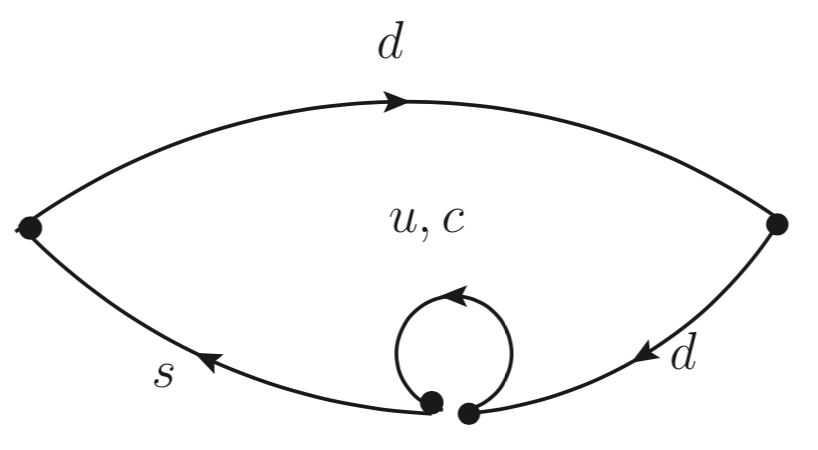}\hfill
        \caption{$K$ to $\pi$ diagrams. The upper two are figure-8 diagrams contracted with operator $Q_1$ (left) and $Q_2$ (right). The lower two are eye diagrams contracted with operator $Q_1$ (left) and $Q_2$ (right). }
        \label{fig:scaling_test_ktopi_diags}
\end{figure}

We can then examine how the matrix elements extracted from three-point functions scale on these lattice ensembles. The three-point diagrams which contribute to $\langle \pi | Q_i | K^0 \rangle$ matrix elements are shown in Figure \ref{fig:scaling_test_ktopi_diags}. Compared to the eye diagrams shown in the bottom of Figure \ref{fig:scaling_test_ktopi_diags} having self-loop parts, the figure-8 diagrams shown in the top of Figures \ref{fig:scaling_test_ktopi_diags} have relatively small statistical errors and don't involve the heavy charm quark which is the most probable source of large discretization error.



If we can compare the results for only the contribution of the figure-8 diagrams to these three-point functions, we can test the scaling violation with high precision. To perform such scaling violation tests on the three-point functions and also on the four-point functions which we will present later, we have to establish that the set of diagrams we are studying is a well-defined portion of the full physical amplitude by itself and has a continuum limit as the lattice spacing $a$ approaches 0. In fact such a continuum limit can be established based on a lower-level understanding of renormalization and the continuum limit.  A natural  collection of graphs to study in isolation in a lattice calculation is that in which the fermion propagators have a fixed topology, such as the present case of these figure-8 diagrams.  For a lattice calculation with a specific combination of quark propagators the path integral provides a sum over all possible gluon emissions, gluon self-interactions and closed fermion loop insertions.  For such a single quark propagator topology, a continuum limit with $ca^2$ scaling follows from the renormalizability and chiral symmetry of DWF QCD provided the quark propagator topology does not create new divergent sub-diagrams, not present in QCD.  If new divergent sub-diagrams do appear, such as the vertex correction arising from the exchange of a gluon between two of the legs of a four-quark vertex resulting from an insertion of $H_W$, then a continuum limit with $ca^2$ scaling will still be guaranteed if these same graphs appear in the NPR subtractions that are performed, a consideration which determines the quark propagator topology used for the NPR procedure. Therefore, we would expect that the difference between such a figure-8 diagram and its continuum limit can be described by $c  a^2$ where $c$ is approximately a constant and the possible logarithmic corrections to the $c  a^2$ behavior are neglected.

\begin{table}[!htbp]
\resizebox{\textwidth}{!}{
\begin{tabular}{|c|c|cc||cc|c|} \hline
     &            &         \multicolumn{2}{c||}{Z factors}          & \multicolumn{2}{c|}{Matrix elements in physical units}    &     \\ 
$\mu$/GeV & Irrep  & 32I         & 24I      & 32I          & 24I &  Scaling violation      \\
     &        & ($a^{-1}=$2.38GeV)   & ($a^{-1}=$1.78GeV)   & ($a^{-1}=$2.38GeV) & ($a^{-1}=$1.78GeV) & \\ \hline
2.15 & (84,1) & 0.52997(11) & 0.47143(8)  & 0.003957(18) & 0.004045(18) & -2.19 \%\\
     & (20,1) & 0.58755(14) & 0.57493(26) & 0.011949(65) & 0.009936(59) & 18.39 \%\\  \hline
2.64 & (84,1) & 0.52489(6)  & 0.46996(6)  & 0.003919(18) & 0.004032(18) & -2.84 \%\\
     & (20,1) & 0.60358(11) & 0.58239(11) & 0.012275(67) & 0.010065(60) & 19.78 \%\\  \hline
\end{tabular}
}
\caption{The Z factors of NPR in $(\gamma_{\mu},\gamma_{\mu})$ scheme and $\langle \pi | Q_{\pm} | K^0 \rangle$(figure-8 only) in physical units on the two lattice ensembles and different scale $\mu$. The relative scaling violations are listed in the last column.}

\label{tab:npr_gamma_multimu_a}
\end{table}
\begin{table}[!htbp]
\resizebox{\textwidth}{!}{
\begin{tabular}{|c|c|cc||cc|c|} \hline
     &            &         \multicolumn{2}{c||}{Z factors}          & \multicolumn{2}{c|}{Matrix elements in physical Unit}    &     \\ 
$\mu$/GeV & Irrep  & 32I         & 24I      & 32I          & 24I &  Scaling violation      \\
     &        & ($a^{-1}=$2.38GeV)   & ($a^{-1}=$1.78GeV)   & ($a^{-1}=$2.38GeV) & ($a^{-1}=$1.78GeV) & \\ \hline
2.15 & (84,1) & 0.60490(35) & 0.55073(40) & 0.004516(21) & 0.004725(22) & -4.51\%\\
     & (20,1) & 0.67062(61) & 0.67164(46) & 0.013638(74) & 0.011608(69) & 16.08\%\\ \hline
2.64 & (84,1) & 0.58968(16) & 0.53025(13) & 0.004403(20) & 0.004549(21) & -3.27\%\\
     & (20,1) & 0.67807(31) & 0.65711(32) & 0.013790(75) & 0.011357(68) & 19.35\%\\ \hline
\end{tabular}
}
\caption{The Z factors of NPR in $(\gamma_{\mu},\slashed{q})$ scheme and $\langle \pi | Q_{\pm} | K^0 \rangle$(figure-8 only) in physical units on the two lattice ensembles and different scale $\mu$. The relative scaling violations are listed in the last column.}
\label{tab:npr_qslash_multimu_a}
\end{table} 

Based on the relationship $Q_{\pm}= (Q_1 \pm Q_2)$, we can easily obtain the matrix elements $\langle \pi | Q_{\pm} | K^0 \rangle$ from linear combinations of results from $Q_i$ operators. The results for these figure-8 diagrams are listed in Table \ref{tab:npr_gamma_multimu_a} and Table \ref{tab:npr_qslash_multimu_a} at $\mu = 2.15$ GeV and $\mu = 2.64$ GeV. The figure-8 matrix element of the operator $Q_+$ which belongs to the (84,1) representation has a small scaling violation of size $\sim 2-4\%$, while the figure-8 matrix element of the operator $Q_-$ which belongs to the (20,1) representation has a large scaling violation of size $\sim 20\%$. 

Even in the absence of a heavy charm quarks, such an unexpectedly large scaling violation as appears in the matrix element of $Q_-$ operator is not unique. As shown in our previously published paper\cite{K2pipi_DeltaI_3_2_PhysRevD.91.074502}, $K\rightarrow \pi \pi$ matrix elements calculated from operators belonging to the (8,8) irreducible representation also show similarly large finite lattice spacing errors as shown in Table XIV of Reference  \cite{K2pipi_DeltaI_3_2_PhysRevD.91.074502}.

\subsection{Scaling of four-point single-integrated correlation functions}
\label{sec:scaling_4pt_tp12_only}
Similar to the three-point scaling tests, we perform a series of scaling tests for the contribution from four-point diagrams of type 1 and type 2, which are all connected. We also need to calculate three-point matrix elements $ \langle \pi | Q_{\pm}^{lat} | K^0 \rangle$ to remove the exponentially increasing terms from the single-integrated four-point correlators.  

In this case, only connected diagrams are calculated, and only up quark can appear in our intermediate states. When we calculate the three-point matrix elements $ \langle \pi | Q_{\pm}^{lat} | K^0 \rangle$, we must use the interpolating operator $O_{\pi^0}=i\overline{u}\gamma_5 u$ rather than  $O_{\pi^0}=i(\overline{u}\gamma_5 u - \overline{d}\gamma_5 d )/\sqrt{2} $ and only include figure-8 diagrams shown in Figure \ref{fig:scaling_test_ktopi_diags} since without disconnected diagrams the combination $\overline{u}\gamma_5 u$ and $\overline{d}\gamma_5 d$ behave as independent degenerate mesons\cite{Jianglei_klks_PhysRevD.88.014508}.

\begin{figure}[htbp]
            \centering
            \includegraphics[width=0.49\textwidth]{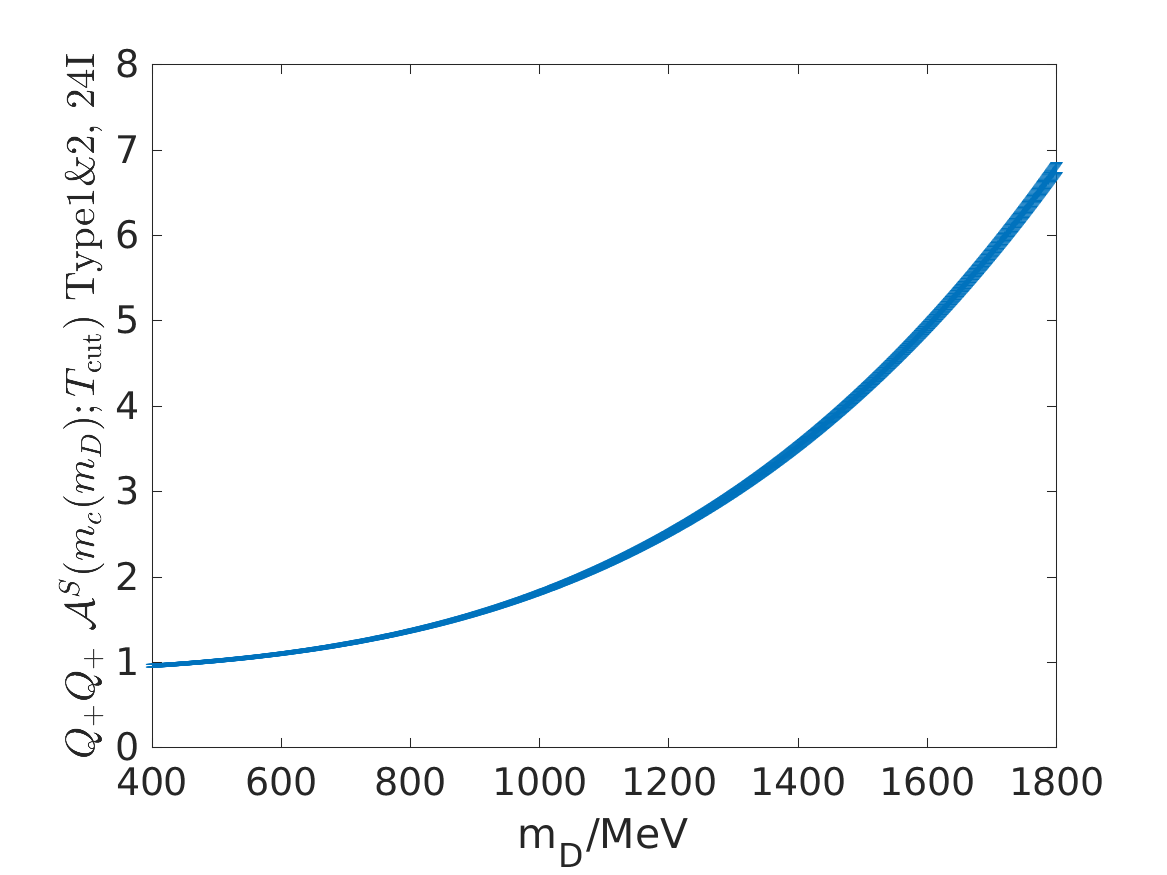} 
            \includegraphics[width=0.49\textwidth]{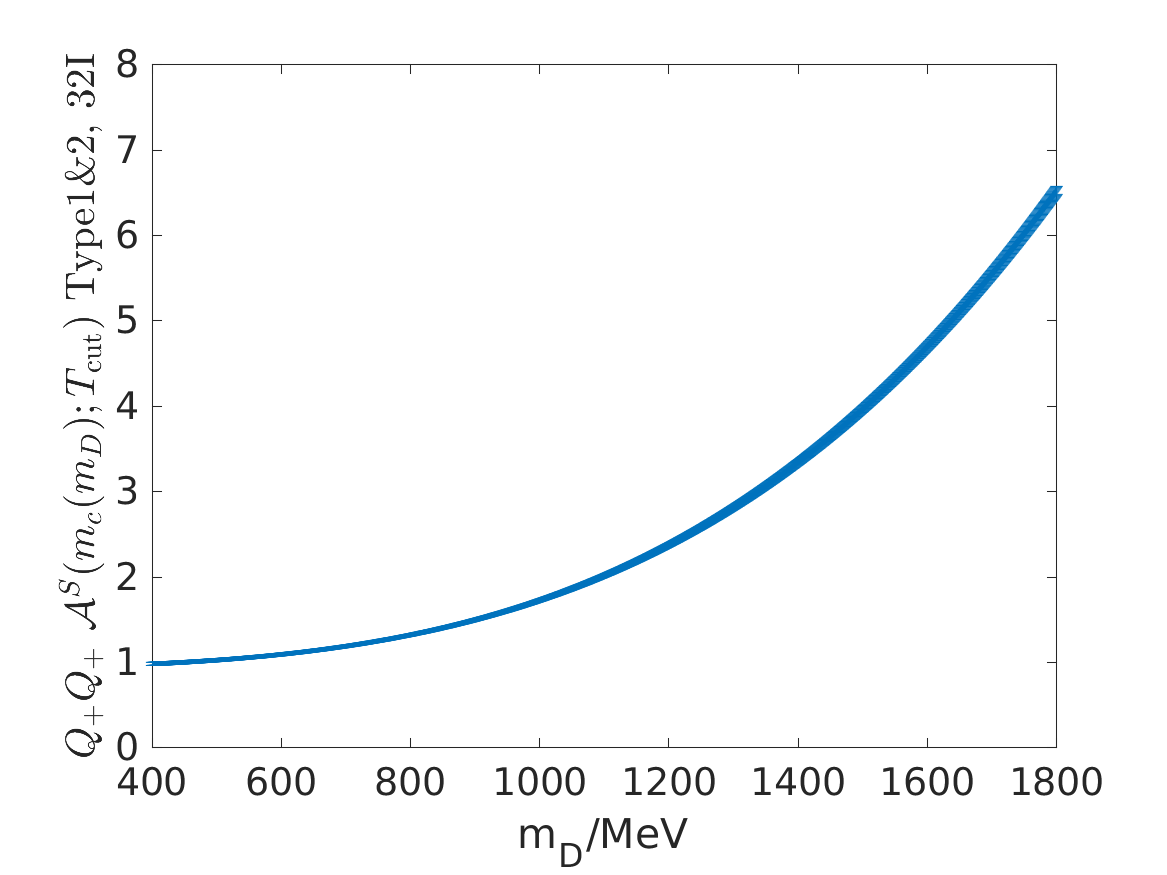}
            \caption{The single-integrated correlators with two $Q_+$ operators plotted as a function of $m_D$ on the 24I ensemble (left) and on the 32I ensemble (right). }
            \label{fig:Qplus_single_int_mc_mD_24_32}
            \end{figure}
    \begin{figure}[htbp]
            \centering
            \includegraphics[width=0.65\textwidth]{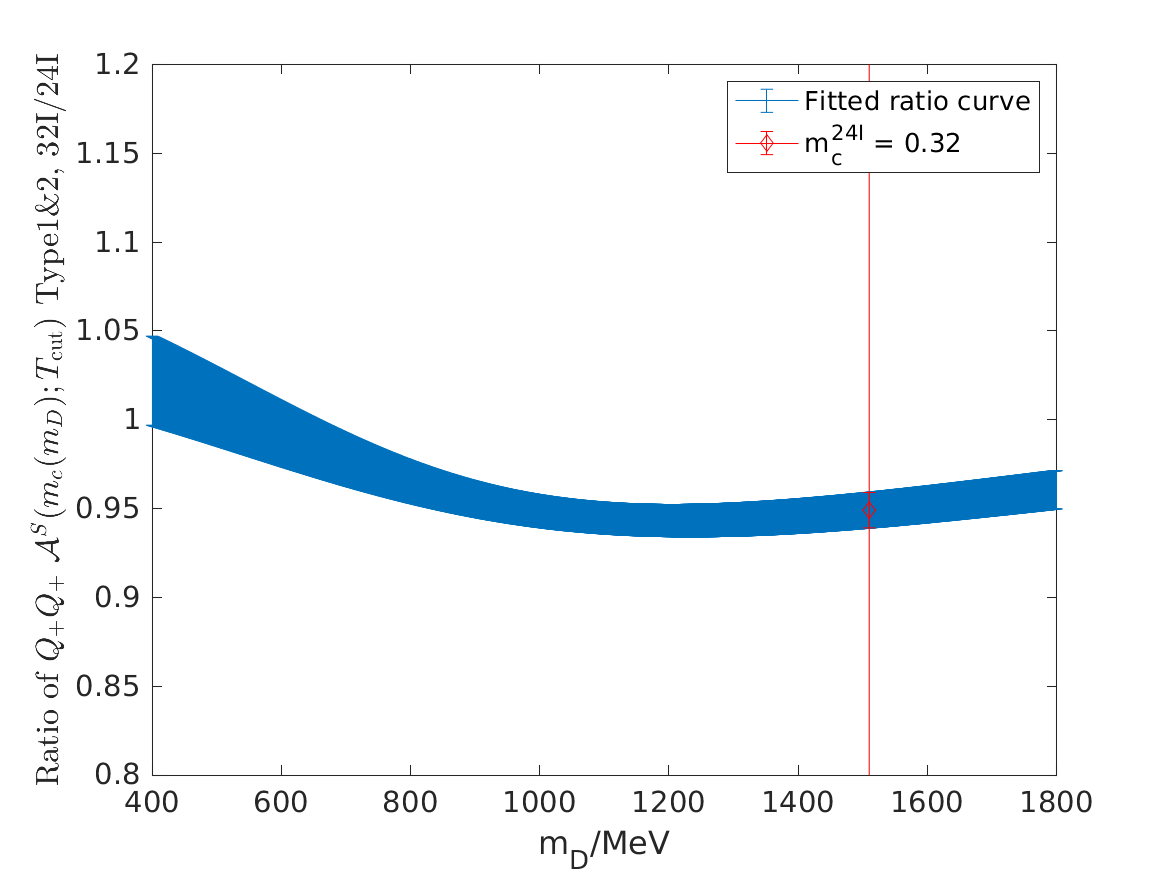} 
            \caption{The ratio of the single-integrated correlators with two $Q_+$ operators on the 32I and 24I ensembles plotted as a function of $m_D$. }
            \label{fig:Qplus_ratio_single_int_mc_mD_32_24}
    \end{figure}

We perform the scaling tests on the single-integrated four-point correlation functions. For the relatively light input charm masses used here, the correlation function is highly non-local and limited by the lattice size, we can not use a sufficiently large $T$ to extract $\Delta m_K$ from the single integration as discussed in Section \ref{sec:dmk_formalism}. However, because the single-integrated correlator itself is a physical quantity with a continuum limit,  we can perform the scaling tests on the single-integrated four-point correlators for the operators $Q_{\pm}$ if we use consistent physical integration ranges on the two different lattices.
%
%
To remove the $O(a)$ errors from difference in the integration range for the single integration, we perform interpolations on the 24I ensemble to match the integration cutoff value on the 32I ensemble, which is $T_{\mathrm{cut}} = 5.87$ GeV$^{-1}$ and evaluate the integral using the trapezoidal rule.

The single-integrated correlators with two $Q_+$ operators and a fixed integration cutoff $T_{\mathrm{cut}} = 5.87$ GeV$^{-1}$ are plotted as a function of D meson mass in Figure \ref{fig:Qplus_single_int_mc_mD_24_32}.
We take the ratios between the results on the two lattices with the same D meson masses in physical units for various charm masses. If the charm quark is the dominating source of scaling violation, as we reduce the charm mass, the ratio between different lattice spacing should approach 1. This can be verified in Figure \ref{fig:Qplus_ratio_single_int_mc_mD_32_24}. On the 64I ensemble, the $m_ca=0.32 \sim 0.33$ gives the physical D meson masses. To estimate the finite lattice spacing effect for our lattice calculation on the 64I ensemble, we mark the point where $m_ca=0.32$ on the coarser 24I ensemble and find the scaling violation is about 5\%.
\begin{figure}[hbt!]
            \centering
            \includegraphics[width=0.5\textwidth]{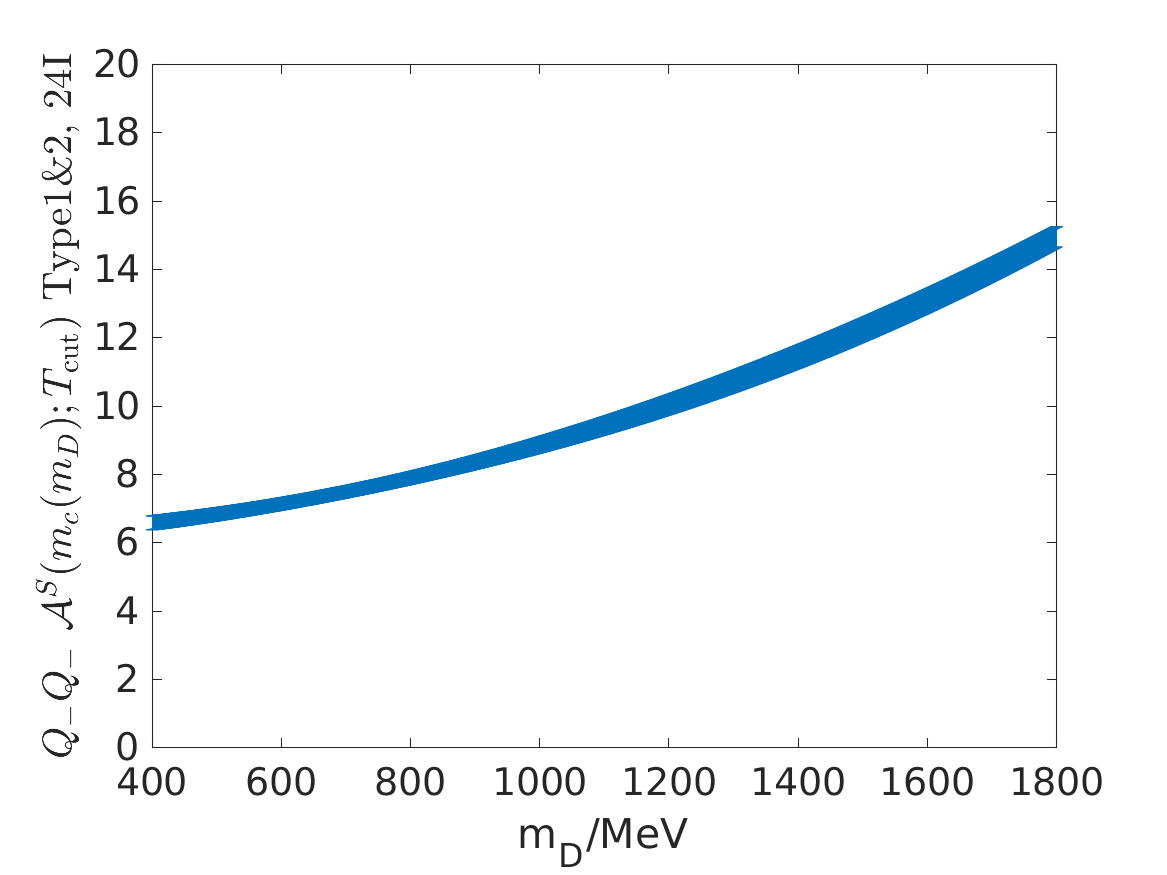}\hfill 
            \includegraphics[width=0.5\textwidth]{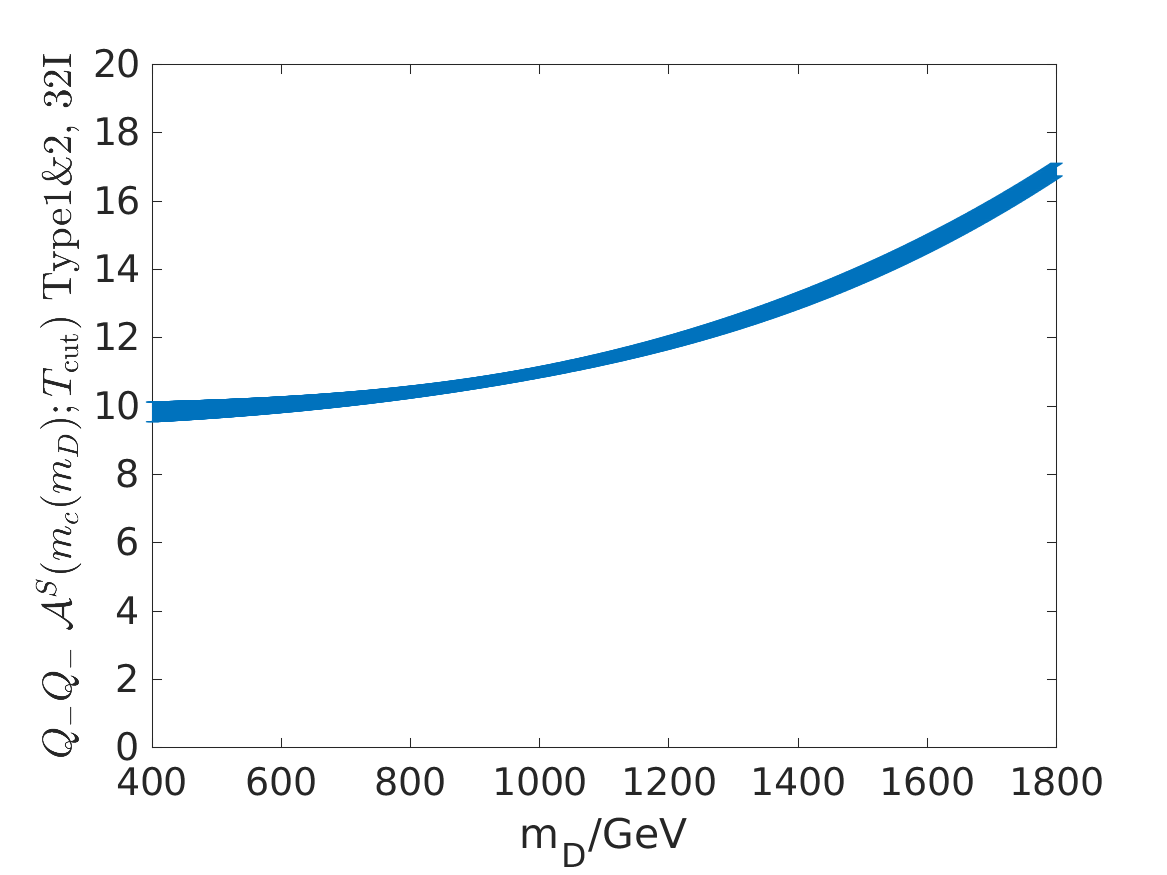} \hfill  
            \caption{The single-integrated correlators with two $Q_-$ operators plotted as a function of $m_D$ on the 24I ensemble (left) and on the 32I ensemble (right). }
            \label{fig:Qminus_single_int_mc_mD_24I_32I}
    \end{figure}
    \begin{figure}[hbt!]
            \centering
            \includegraphics[width=0.65\textwidth]{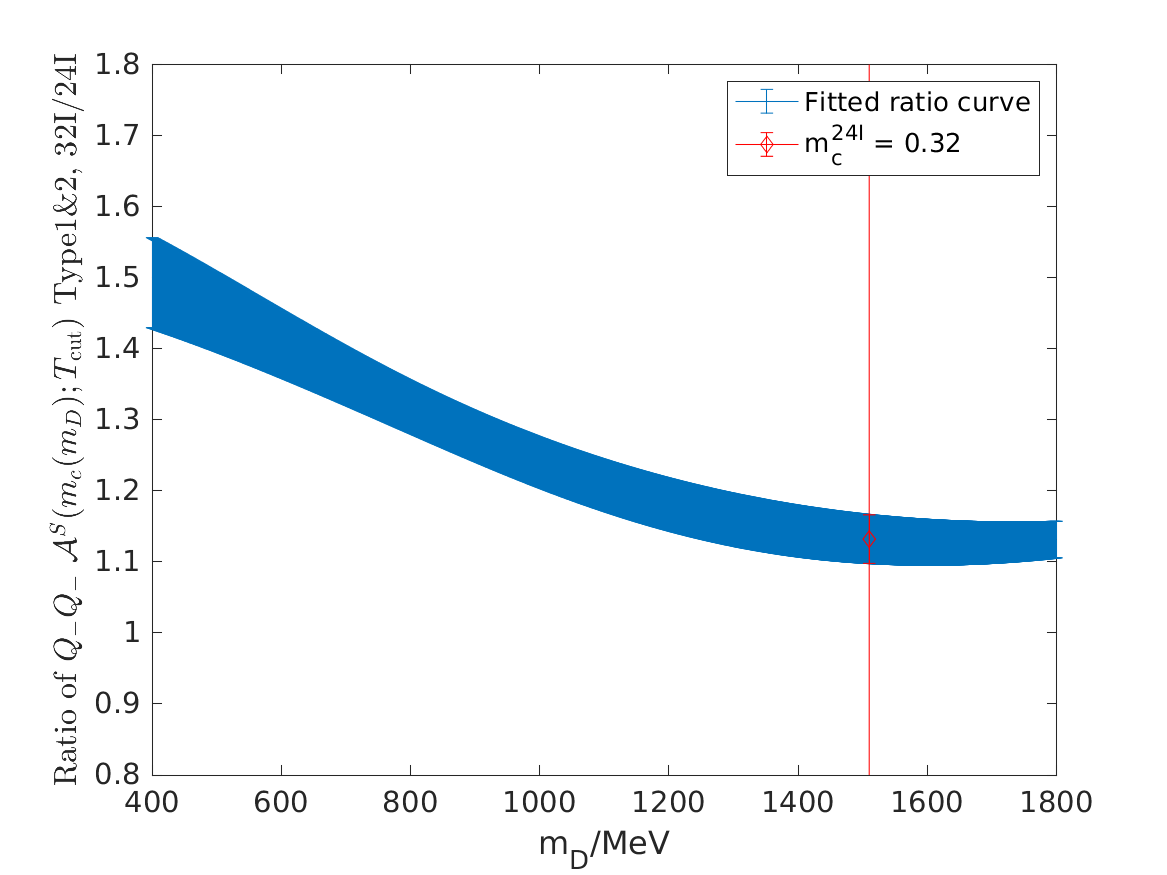} 
            \caption{The ratio of the single-integrated correlators with two $Q_-$ operators on the 32I and 24I ensembles plotted as a function of $m_D$. }
            \label{fig:Qminus_ratio_single_int_mc_mD_32_24}
    \end{figure}

Similarly, 
in Figure \ref{fig:Qminus_single_int_mc_mD_24I_32I}, we plot the single-integrated correlators with two $Q_-$ operators on the 24I and 32I ensembles. In Figure \ref{fig:Qminus_ratio_single_int_mc_mD_32_24}, we plot the ratios between the results on the two lattices as a function of physical D meson mass. The scaling violation at $m_ca=0.32$ is about 14\%. However, we find the ratio for the case with two $Q_-$ operators is not approaching 1 as the charm mass becomes smaller but instead approaching a ratio which is about 1.4. This indicates that in addition to the scaling violation introduced by the heavy charm quark, the scaling error for the four-point integrated correlators with two $Q_-$ operators, can be as large as 40\%. 

In our $\Delta m_K$ calculation, we have combinations of four-point correlation functions with $Q_+$ and $Q_-$ operators. Based on the scaling tests performed on the 24I and 32I ensembles, we estimate the finite lattice spacing error to be of order of $40\%$.
\section{Conclusion and outlook}     
Our preliminary result for $\Delta m_K$ based on 152 configurations with physical quark masses is:
        \begin{equation}
            \Delta m_K  = 5.8(0.6)_{\mathrm{stat}}(2.3)_{\mathrm{sys}} \times 10^{-12} \mathrm{MeV}.
        \end{equation}
Here the first error is statistical and the second is an estimate of largest systematic error, the discretization error, based on the scaling tests performed on the 24I and 32I ensembles. A comparison between our $\Delta m_K$ value and the experimental value $3.484(6) \times 10^{-12}$ MeV \cite{PDG_2020} suggests reasonable agreement given the large finite lattice spacing errors. Future calculations on a $96^3 \times 196$ lattice together with this completed calculation on the $64^3 \times 128$ ensemble with physical quark masses, will allow the continuum limit to be explored.


\printbibliography

\end{document}